\documentstyle[aps,preprint]{revtex}
\tighten
\begin{document}
\draft
\title{\bf Pulsed Magnetic Field Measurements of the
Composite Fermion Effective Mass}
\author{D.R.~Leadley,$^1$ M.~van~der~Burgt,$^1$ R.J.~Nicholas,$^1$
C.T.~Foxon$^2$ and
J.J.~Harris$^3$}
\address{$^1$Department of Physics, Oxford University, Clarendon
Laboratory, Parks Road, Oxford, OX1~3PU,~UK\\
$^2$Department of Physics, Nottingham University,
University Park, Nottingham, NG7~2RD,~UK\\
$^3$Department of Electronic Engineering, University College
London.~UK}

\date{Submitted to Phys. Rev. B  May 11, 1995}
\maketitle

\begin{abstract}
Magnetotransport measurements of Composite Fermions (CF) are reported in 50 T
pulsed magnetic
fields. The CF effective mass is found to increase approximately linearly with
the effective field
$B^*$, in agreement with our earlier work at lower fields. For a $B^*$ of 14 T
it reaches
$1.6m_e$, over 20 times the band edge electron mass. Data from all fractions
are unified by the
single parameter $B^*$ for all the samples studied over a wide range of
electron densities. The
energy gap is found to increase like $\sqrt{B^*}$ at high fields.
\end{abstract}
\pacs{73.40.Hm, 73.20.Dx, 72.20.Jv}

\narrowtext
The Fractional Quantum Hall Effect (FQHE) has been known for many
years,\cite{tsui} but a
complete understanding has remained elusive due to the complex
electron-electron interactions
responsible. Recently a great deal of excitement was aroused by a model in
which the electrons are
transformed into Composite Fermions (CF).\cite{jain,hlr,cfexpts,lead,du,mano}
In this model the
Coulomb interaction of one electron with all the others is replaced with a
Chern-Simons gauge field,
equivalent to attaching an even number ($2m$) of flux quanta ($\Phi_0=h/e$) to
each electron. In a
mean field approximation the gauge field exactly balances the external field at
filling factor $\nu =
1/2m$ where the system of interacting electrons in high magnetic field is
replaced by one of
independent CFs in zero field. At other filling factors there are more (or
less) flux quanta than
required to cancel the gauge field and the CFs see an effective magnetic field,
$B^*=B-
2m\Phi_0n_e$. This leads to quantisation of the CF energy into Landau levels
(LLs) and gaps open
in exact analogy with the Integer Quantum Hall Effect (IQHE) of non-interacting
electrons. Thus the
FQHE may be simply regarded as the IQHE of Composite Fermions.

Several experiments have shown that CFs appear to behave like real
particles,\cite{cfexpts} in that
they have a well defined Fermi wavevector and follow particle like trajectories
under the influence of
the effective magnetic field. Therefore it is of interest to know if an
effective mass may be associated
with these objects and what the origins of such a mass are. In an earlier work
\cite{lead} (hereafter
{\bf I}) we treated the FQHE features in $\rho_{xx}$ as Shubnikov-de~Haas
oscillations (SdHO)
and analysed the temperature dependence of the amplitudes to determine the CF
effective mass
$M^*$ for three samples of low electron density $n_e$. In this paper we report
similar
measurements for higher density samples in pulsed magnetic fields up to 50T,
and observe a strong
field dependence of $M^*$ which is related to the electron-electron
interactions. This report also
presents both the highest electron density and the highest magnetic field at
which the FQHE has been
studied.

In our earlier work $M^*$ was evaluated for each maximum and minimum between
$\nu=2/3$ and
1/3, and found to be given by $M^*=(0.51\pm0.05)+(0.074\pm0.015)B^*$, in units
of the free
electron mass $m_e$. The same dependence was found for positive and negative
values of $B^*$,
i.e.\ for fractions above and below $\nu=1/2$, and the mass values from samples
with $n_e$
differing by a factor of 2 was the same at a given value of $B^*$. This
suggested that $B^*$, as
opposed to the externally applied field $B$, was the important parameter and
that the CF model
was appropriate. For the samples studied in {\bf I}, with $n_e$ between 0.6 and
1.2$\times
10^{15}$m$^{-2}$, the maximum value of $B^*$ accessed was 3.7T for the
$\nu=1/3$ minimum,
at which point $M^*=0.78m_e$. This is some 40\% larger than measured at the
lowest effective
field of 0.5T for 5/9. The experimental errors arising from uncertainty in the
electron temperature and
the fitting procedure are $\sim0.1m_e$, however by now performing the
experiment at higher
effective fields we can test the dependence of $M^*$ on $B^*$ to a much greater
degree. There
has been much speculation as to what the mass does as $\nu=1/2$ is approached,
where the
oscillations in $\rho_{xx}$ are weak, with some reports of a strong
divergence.\cite{du} Our
current data does not support this divergence, but in the present paper we
restrict our analysis to
regions where the oscillations in $\rho_{xx}$ are reasonably large at low
temperatures.

The measurements were performed on Hall bars of high quality GaAs-GaAlAs
heterojunctions
grown by MBE at Philips Research Laboratories.\cite{jeff} Three samples (G627,
G650, and
G902) had significantly lower impurity contents due to the use of a short
period AlAs/GaAs
superlattice in the buffer layer. The samples were measured after
photoexcitation to ensure maximum
homogeneity and the largest possible features in the resistivity. The relevant
parameters of these
samples as measured are given in Table 1 together with those for the samples
reported in {\bf I} for
completeness.

The resistivity $\rho_{xx}$ was measured for temperatures between 550mK and
4.2K using a
$^3$He cryostat in pulsed magnetic fields of up to 50T. The temperature was
measured with a
ruthenium oxide resistor mounted next to the sample. Good thermal equilibrium
was ensured by
keeping both the resistor and the sample immersed in $^3$He liquid throughout
the pulse. By
comparing data obtained in the lower field section of the pulses, on both up
and down sweeps, with
data from the same samples studied in steady fields up to 16T, we estimate that
the temperature rise
during the pulse was less than 50mK. The current used was verified to be
sufficiently low to prevent
electron heating by comparing the peak amplitudes from shots taken with a range
of currents.
Induced voltages from the pulse were minimised by careful attention to the
sample wiring and
eliminated by averaging otherwise identical shots with opposite field
directions. There is a slight shift
in position of features between the two halves of the pulse due to an RC time
constant which can be
corrected for. Data taken as the field decays is generally better since the
field is changing less rapidly
and therefore we concentrate on this in the remainder of the paper.

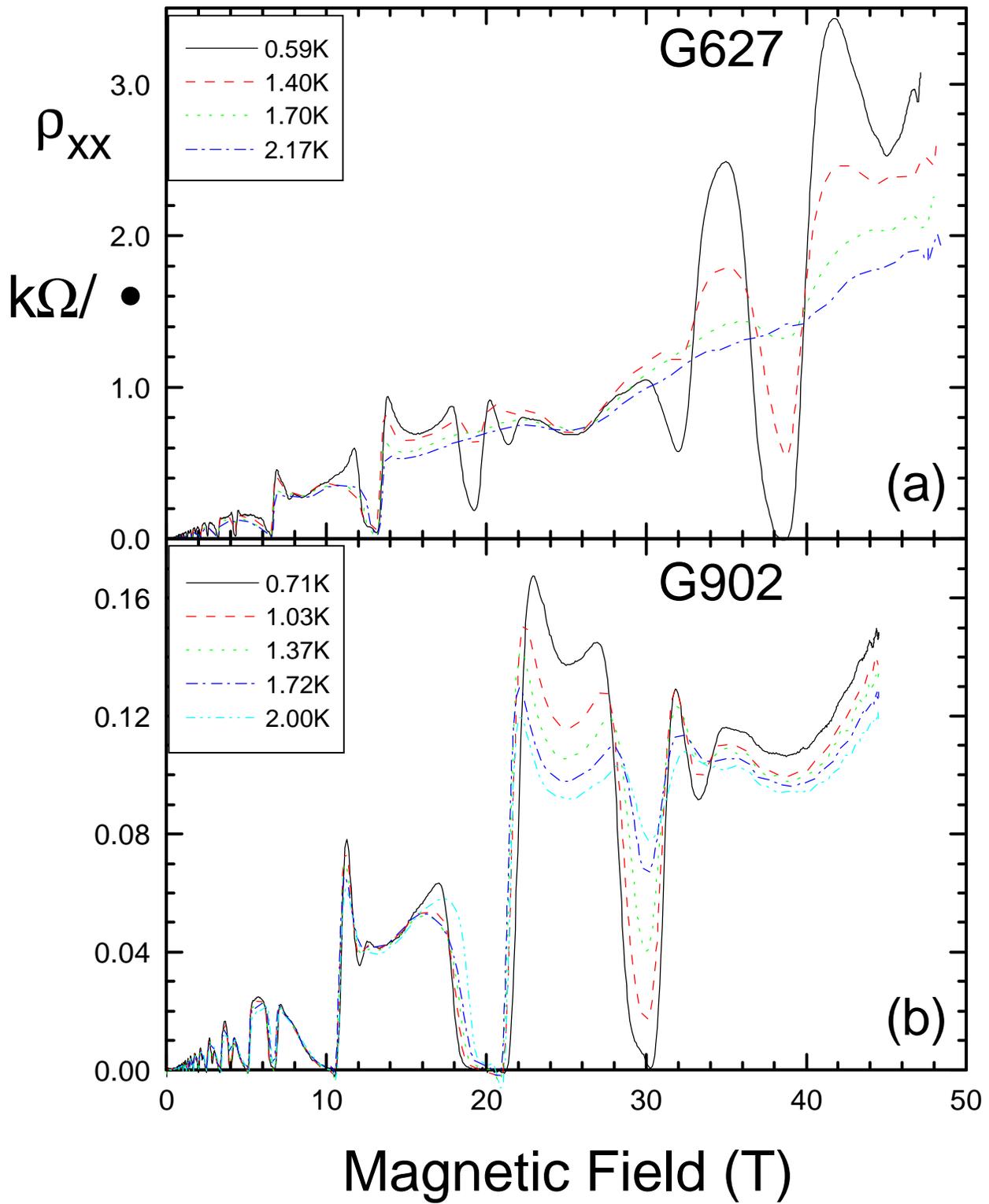
\begin{figure}
\caption{Resistivity of (a) sample G627, and (b) G902b, at temperatures between
0.6 and 2.2K,
showing FQHE features at the highest fields yet observed.}
\label{fig-sdh}
\end{figure}

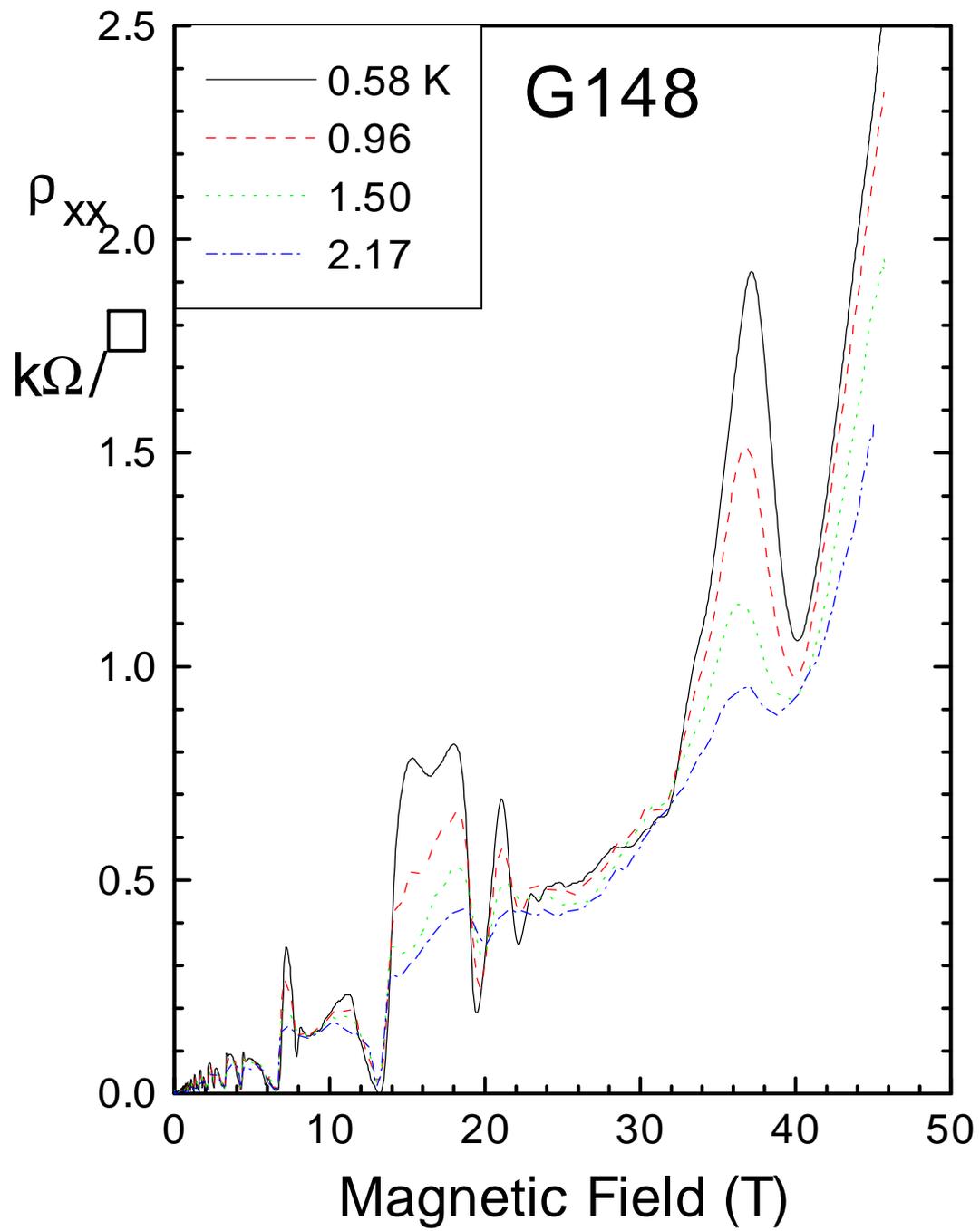
\begin{figure}
\caption{Resistivity of sample G148, showing a rapid increase in resistivity
above 32T at low
temperatures.}
\label{fig-148}
\end{figure}

Figure~\ref{fig-sdh} shows the temperature evolution of the FQHE features for
two of the samples
studied. For sample G627, in Fig.~\ref{fig-sdh}(a), we see $\nu=1/3$ at 38T and
a well developed
2/7 fraction at 45T, which is the highest field at which the FQHE has yet been
observed. In
Fig.~\ref{fig-sdh}(b) data is presented from the highest electron density
sample yet studied, G902b
with $\nu=2/3$ at 30T. Taken together with our other data these figures show
that FQHE states
survive to very high magnetic fields in much the same form as at lower fields,
but can now be
observed at much higher temperatures. Similar data was obtained for samples
G650 and G902a.
However for sample G148 the resistance increased rapidly above 32T at low
temperature as can be
seen in Fig.~\ref{fig-148}. This means that of the features at fields above
$\nu=1/2$ only the 1/3
minimum can be reliably analysed, although those below 1/2 are unaffected. The
onset of the rapid
increase is near the 3/7 fraction, but does not appear to be related to the
filling factor. We ascribe
this increase of resistance to magnetic field induced localisation since there
is a greater background
impurity concentration in sample G148. Unlike the G600 series of samples, it
does not have a short
period superlattice in the GaAs buffer layer. Similar dramatic increases in
resistance have been seen
around 1/3 in GaAs hole gases, also having a large impurity content, which was
ascribed to
magnetically induced Wigner crystallisation.\cite{holes} However, in 2DEGs
Wigner crystallisation is
not thought to occur at filling factors greater than 1/3 and so we do not
believe these increases in
resistance are due to any periodic arrangement of the electrons.

The oscillations in resistivity are described by the Ando formula\cite{ando}
\begin{equation}
\frac{\Delta\rho_{xx}}{\rho} \propto \frac{X}{\sinh X}\:
\exp\left(-\frac{\pi}{\omega_c\tau_q}\right)\:\cos2\pi(\nu-1/2),
\label{eqn-ando}
\end{equation}
\noindent where $X=2\pi^{2}k_{B}T/\hbar\omega_{c}$ and $\omega_{c}=eB/m^*$ is
the
cyclotron frequency. For Composite Fermions we replace $B$ by $B^*$, $\nu$ by
$\nu^*$,
$m^*$ by $M^*$ and $\tau_q$ by ${\cal T}_q$. From the temperature dependence of
the
oscillations a value of $M^*$ can be obtained at each maximum and minimum,
using the amplitude
of the oscillation envelope and the effective field at which the extrema occur.
Since Eq.~(\ref{eqn-ando}) is valid for all fields, not just the FQHE minima, a
value for $M^*$ can
be obtained at any field provided that data from the same measured $B^*$ is
used at all
temperatures. It is necessary to assume that ${\cal T}_q$ (i.e.\ the level
broadening) does not
change with temperature over the region of interest. In the case of GaAs-GaAlAs
electron gases at
high fields we believe this to be a justifiable assumption. However, in cases
where ${\cal T}_q$ is
not constant any mass values obtained must be treated with suspicion.

The analysis was performed on the FQHE features that are large at low
temperature, i.e.\ we do not
consider the weak features near to $\nu=1/2$ where there are extremely large
experimental
uncertainties in the size of the oscillations and only a very limited
temperature range where they can
be distinguished from background noise. For each feature that we choose to
analyse, only data for
which $\Delta\rho/\rho<50\%$ are considered, thus the FQHE features are always
only a weak
modulation of the conductivity and appear predominantly as sinusoidal
oscillations periodic in
$1/B^*$. For a feature at a given magnetic field this condition can always be
satisfied by choosing a
suitable temperature range. At lower temperatures the oscillations are less
sinusoidal with the minima
approaching zero resistivity and the positions of the maxima especially
shifting from the expected
field positions. Since the low temperature traces have the largest features the
initial impression gained
from Fig.~\ref{fig-sdh} may be that these FQHE oscillations are not sinusoidal,
but it should be
remembered firstly that we are very careful only to analyse small oscillations
and secondly that the
usual electron SdHO show just the same behaviour. Higher harmonics can be
included in
Eq.~(\ref{eqn-ando}) and we indeed find that $aX/\sinh X + b(X/\sinh X)^2$ fits
both the electron
and CF SdHO accurately up to $\Delta\rho/\rho \sim90\%$, but that for small
oscillations the extra
term does not significantly alter the value of X deduced. Therefore to reduce
the number of fitting
parameters we have not pursued this approach in the analysis of this paper,
choosing instead to limit
the range of data considered.
The derivation of Eq.(\ref{eqn-ando}) assumes a constant background resistance
i.e.
$\rho(B)=\rho(0)$ and shows how large a fraction of this background the
oscillations are. However,
for a sample like G148 where the features occur on a rising background it is
necessary to consider
the ratio $\Delta\rho/\rho(B)$ to obtain a good fit to Eq.(\ref{eqn-ando}), as
opposed to just
analysing the change in $\Delta\rho$ or $\Delta\rho/\rho(B=0)$. The agreement
with data from all
the other samples justifies this approach, which we also used to analyse second
generation CF
features around $\nu=1/4$ in {\bf I}.

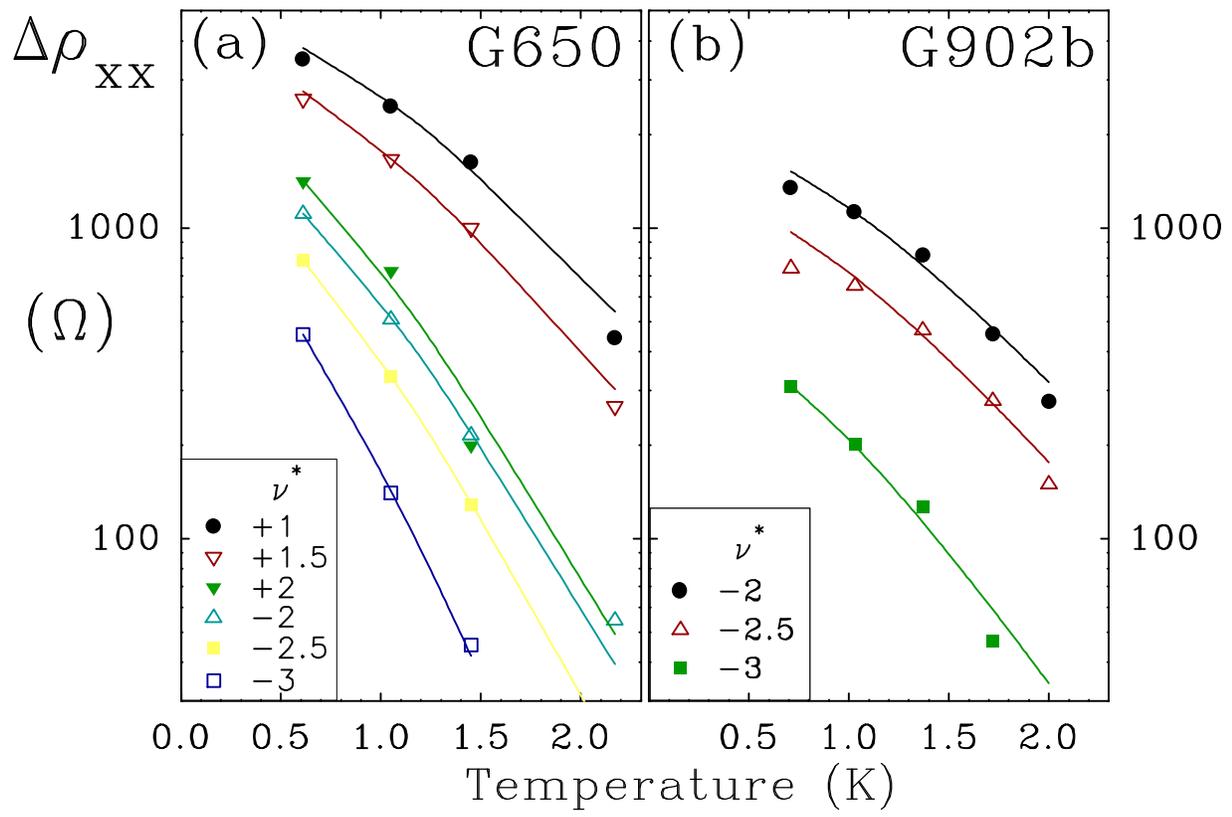
\begin{figure}
\caption{Temperature dependence of the resistivity oscillations of (a) sample
G650, and (b) G902b,
with the fits to the Ando formula shown by dashed lines.}
\label{fig-fit}
\end{figure}

Figure~\ref{fig-fit} shows typical measured and fitted amplitudes for samples
G650 and G902b. In
each case the largest features are not included in the least squares fit to
Eq.~(\ref{eqn-ando}) at the lowest temperatures, where there are noticeable
signs of saturation as
the minima approach zero and our criterion is not satisfied. We find that
Eq.~(\ref{eqn-ando}) fits
the data well over more than an order of magnitude change in $\Delta\rho$ and
that it is very
sensitive to the value of $M^*$. Therefore although there may be quite large
errors in measuring the
absolute sizes of the oscillations the experimental uncertainty in $M^*$ is
never more than
$\pm0.2m_e$ and for most cases is less than half this value.

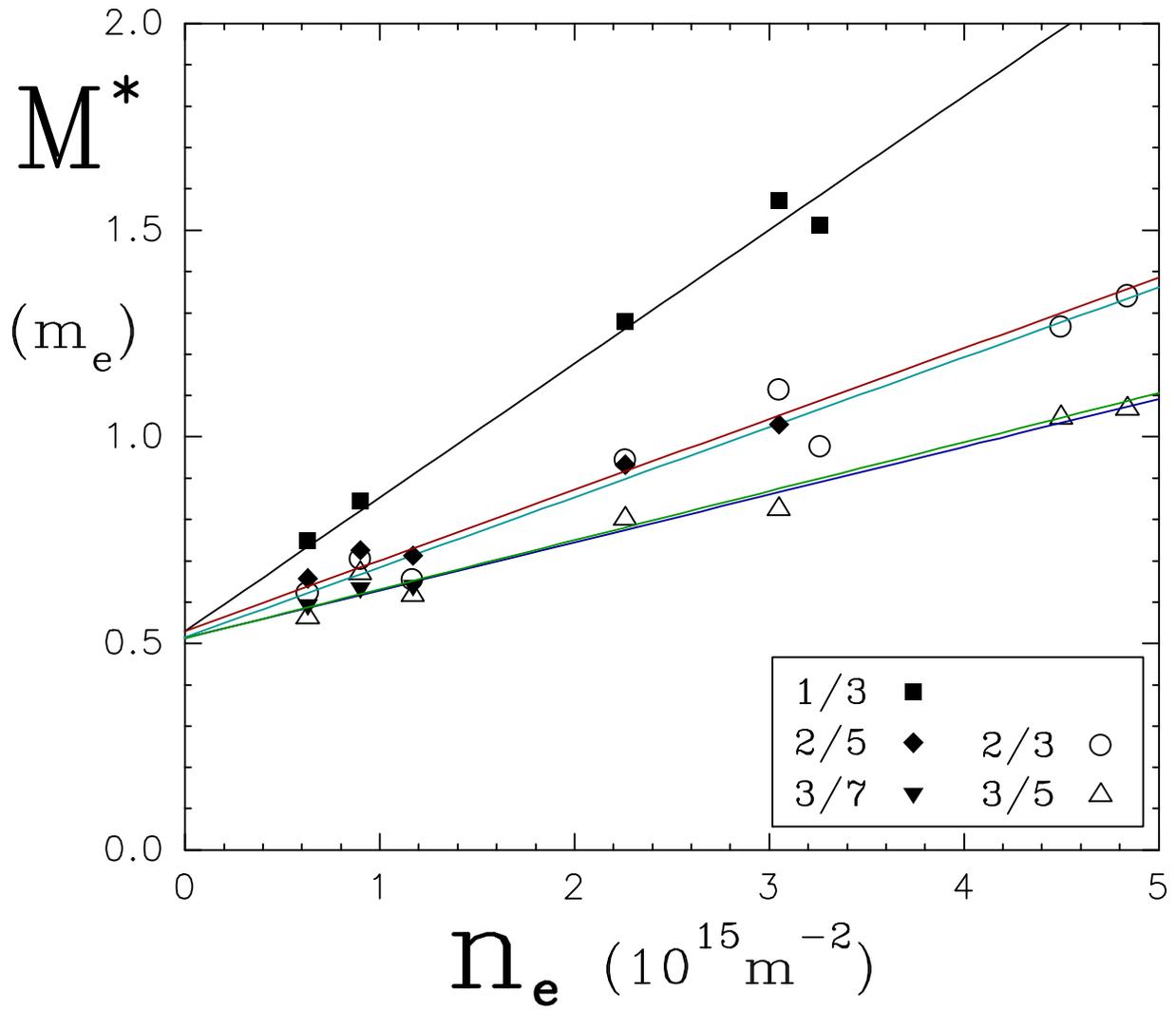
\begin{figure}
\caption{The Composite Fermion effective mass $M^*$ deduced for the maxima and
minima of the
resistivity oscillations shown as a function of carrier density $n_e$ for each
fraction $p/q$. The fitted
lines have gradients $\propto 1/p$.}
\label{fig-massn}
\end{figure}

The results of the analysis are shown in Fig.~\ref{fig-massn} as a function of
$n_e$ for several
different fractions, including data from the samples of {\bf I}. There is no
unique $n_e$ dependence
covering all fractions, but it can be clearly seen that instead the mass values
fall in pairs,
corresponding to states with a common numerator $p$, e.g. 2/3 and 2/5. These
have equal numbers
of occupied CF Landau levels, but occur on either side of $\nu=1/2$ with
effective fields in opposite
senses. This provides a simple demonstration of the symmetry of the states
about $\nu=1/2$ which is
consistent with the CF model, rather than that of particle-hole conjugation
where states of common
denominator $q$ (e.g. 1/3 and 2/3) look similar. In the low density limit all
states tend to the same
effective mass, but the lower index CF Landau levels show an increasing
``non-parabolicity''. By
$n_e=3\times10^{15}$m$^{-2}$ the effective masses for $\nu=1/3$ and $\nu=2/3$
differ by
approximately 40\%, while the masses for $\nu=2/3$ and $\nu=2/5$ differ by less
than 5\%.
Furthermore the gradients of each line on Fig.~\ref{fig-massn} follow
accurately a $1/p$
dependence. The CF mass may then be given by the expression
\begin{equation}
M^* = 0.510 +\frac {0.35}{p} n_e
\end{equation}
in units of $m_e$, with $n_e$ in units of $10^{15}$m$^{-2}$. Since $B^*=\Phi_0
n_e/p$ this
shows there is no additional dependence on $n_e$ above that produced by the
effective field.

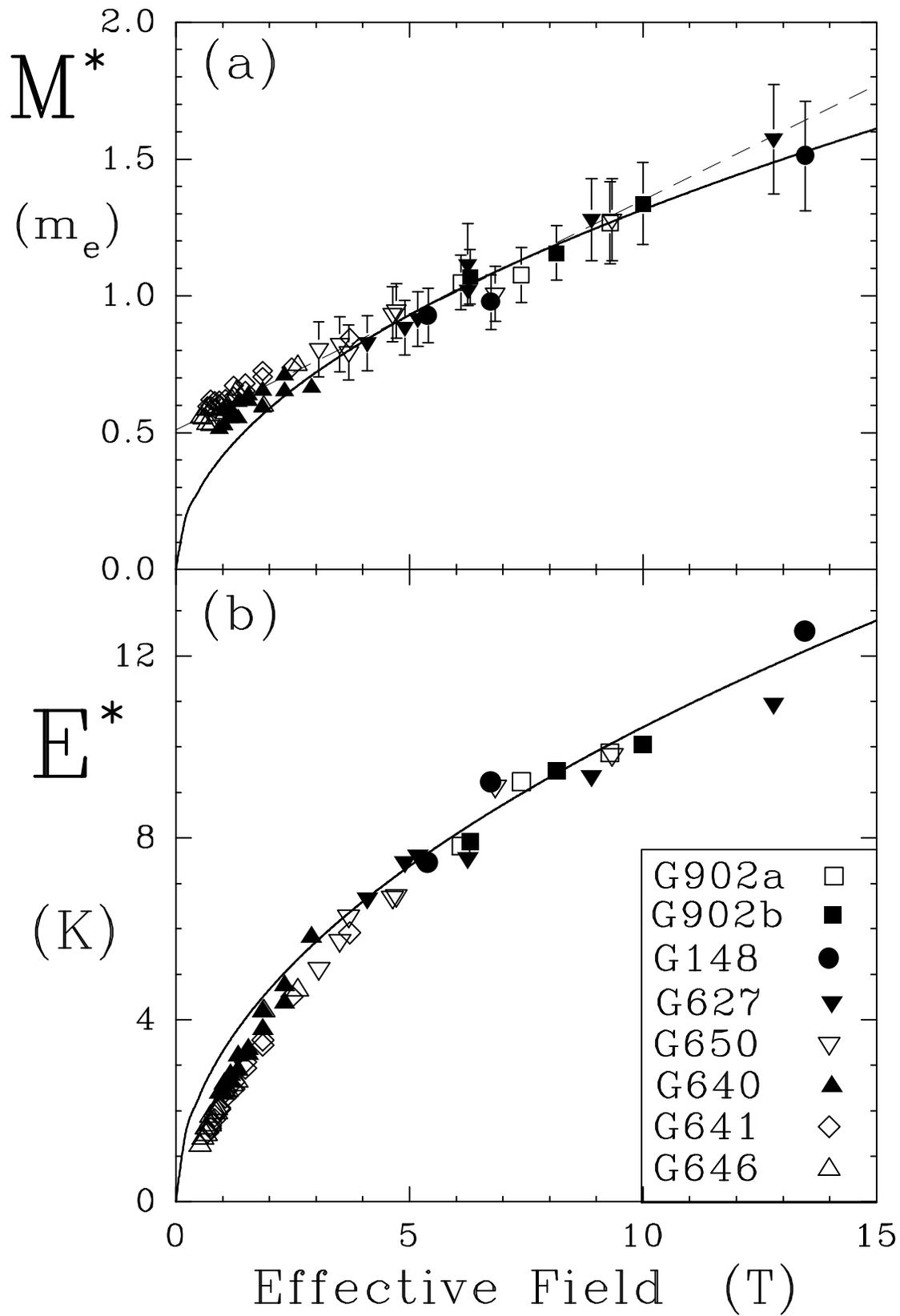
\begin{figure}
\caption{(a) CF effective mass and (b) CF energy gap as a function of the
effective field $B^*$. The
fitted lines are described in the text.}
\label{fig-mne}
\end{figure}

Accordingly, $M^*$ is shown as a function of the effective field in
Fig.~\ref{fig-mne}a, where data
from a large number of different fractions and from samples with densities
varying by an order of
magnitude all lie on a single line. The best fit is shown by the dashed line,
\begin{equation}
M^*=(0.510\pm0.015) + (0.083\pm0.005)|B^*|.
\label{eqn-mass}
\end{equation}
Again in Fig.~\ref{fig-mne}a, as in Fig.~\ref{fig-massn}, there is no
noticeable difference between
data taken from either side of $\nu=1/2$ showing that the sign of $B^*$ is not
important. (In the
remainder of the paper $B^*$ should be read as $|B^*|$). Taken together with
our earlier data, we
can now state that we have shown that there is a simple functional dependence
of the CF mass on
the single parameter $B^*$, covering more than a factor of 25 variation of
$B^*$. At the highest
effective field $M^*=1.6m_e$, which represents an increase of $\sim200\%$ over
our range of
$B^*$, and a factor of 24 enhancement over the GaAs band edge mass.

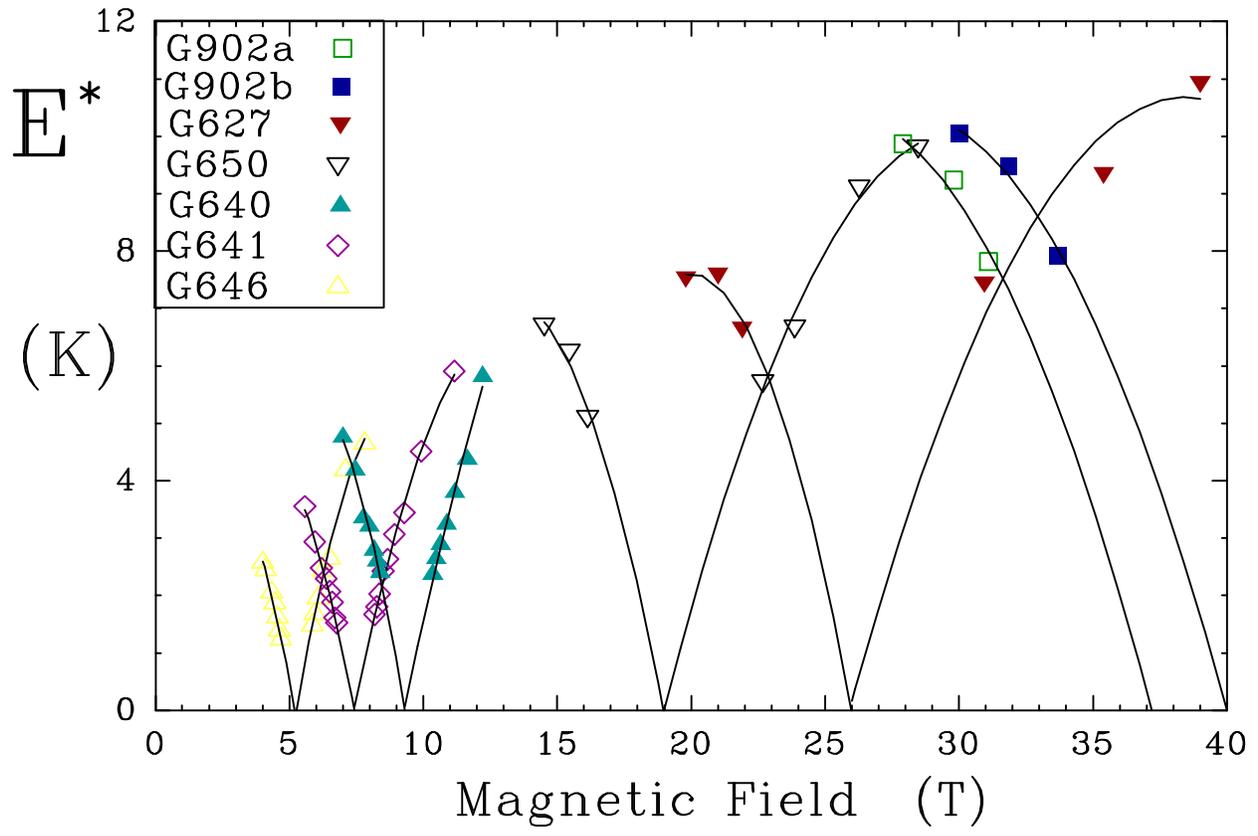
\begin{figure}
\caption{$E^*_c$ plotted as a function of the external field $B$ showing how
this does not lead to
the simple dependence found in Fig.~5b where $B^*$ is the independent variable.
The lines connect
the data to $\nu=1/2$ for each sample.}
\label{fig-mess}
\end{figure}

The measurement of $M^*$ is equally a measurement of the CF cyclotron energy
since
\begin{equation}
E^*_c=\hbar eB^*/M^*,
\label{eqn-e}\end{equation}
i.e.\ $E^*_c$ is the separation of CF LLs. In fact it is the ratio $kT/E^*_c$
(independent of
$B^*$) that determines the oscillations in Eq.~(\ref{eqn-ando}) so we could
regard the energy gap
as the more fundamental quantity, with $M^*$ being a consequence of the model
which considers
CFs in an effective field. This $E^*_c$ is equivalent to what would previously
be known as the
FQHE energy gap $\Delta$, for a sample with infinitely narrow levels. However,
unlike activation
energy measurements which only give the separation between the extended state
regions in adjacent
LLs, this method directly probes the energy difference between the centers of
the LLs and so the LL
width does not need to be known. In Fig.~\ref{fig-mne}b we show $E^*_c$ as a
function of the
effective field for all the samples. With $B^*=14$T we see an energy gap of 12K
for the 1/3
fraction, allowing FQHE features to be seen at quite high temperatures. It is
again remarkable how
plotting $E^*_c$ against the effective field yields a single curve for all
samples and all fractions
considering both maxima, minima and either sense of $B^*$.
By contrast Fig.~\ref{fig-mess}, where the gap is plotted against the real
field, shows no such
orderly behaviour and many additional restrictions have to be added before a
meaningful description
of the data can emerge. (That the data points on Fig.~\ref{fig-mess} do not
extend all the way along
the dotted lines to $\nu=1/2$ for each sample is a consequence of the ranges of
field and
temperature over which the FQHE may be observed.) Thus we make our claim that
the Composite
Fermion effective magnetic field is the single parameter that unifies all the
experimental Fractional
Quantum Hall Effect data.

Following this qualitative conclusion we wish to understand the functional
dependence of $E^*_c$
on $B^*$. The increase of $M^*$ with $B^*$ found in Eq.~(\ref{eqn-mass})
necessarily means
that $E^*_c$ will show a sub-linear increase with effective field, as opposed
to the linear increase
that would be expected of a pure cyclotron energy for particles with a constant
effective mass. The
problem may be visualised in two ways: either (i) $E^*$ is treated as a true CF
cyclotron energy
and we have to explain why the effective mass, as a property of real CF
particles, increases with
$B^*$; or (ii) we explain how $E^*_c$, as the FQHE energy gap, increases with
$B^*$ and treat
$M^*$ as just a parameter defined via Eq.~(\ref{eqn-e}).

Taking the first approach, we recall that the measured single particle electron
effective mass may
increase slightly with electron density, since when the Fermi energy is higher
the conduction band
non-parabolicity becomes important and that this non-parabolicity can be
accurately accounted for
by the proximity of other conduction and valance bands.\cite{nonp} For the
normal low field
SdHOs there is however no change of mass with field since all charge transport
occurs at the Fermi
energy which remains (approximately) constant as it moves between LLs. To
account for the CF
data presented here in terms of non-parabolicity the CF LLs would need to
deviate a lot more from
parabolic bands than is the case for the bare electrons and also the CFs
responsible for conduction
would have to come from much higher in the band as $|B^*|$ increased. Thus
corrections to the
mass from non-parabolicity seems unlikely to be the explanation.

Now we turn to the second case where the energy gap is considered first. The
FQHE is the result of
a many body Coulomb interaction and so theoretically the energy gap has been
written as $\Delta=
C_{\nu} e^2/(4\pi\epsilon\epsilon_0 l_0)$, where $l_0$ is the cyclotron radius
(proportional to the
interparticle spacing) and $C_{\nu}$ is a fixed coefficient, different for each
fraction. Halperin, Lee
and Reed\cite{hlr} (HLR) have used this relationship and argued on dimensional
grounds that the
high field limit of $M^*$ should show a square root dependence on carrier
density through
$l_0$.\cite{hlrnote} This corresponds to a $\sqrt{B}$ dependence for both the
cyclotron energy
and the mass for any given fraction. Since $B^*=B(1-2\nu)$ this approach also
gives a
$\sqrt{B^*}$ dependence, but only for each fraction separately. However our
data clearly shows
that although there is not a single functional dependence on $n_e$ (and hence
$B$) covering all
fractions, if instead we use $B^*$ there is a single dependence for all
fractions and all densities. In
this spirit the data of Fig.~\ref{fig-mne}b has been fitted to
$E^*_c=a\sqrt{B^*}$ as shown by the
dashed line with $a=3.3(\pm0.2)$KT$^{-1/2}$. If the exponent of $B^*$ is also
allowed to vary
the best fit is achieved for a power of 0.57, although if only the high field
data ($B^*>2$T) is
considered the result is 0.51. Thus it seems that a $\sqrt{B^*}$ behavior
adequately describes our
data for the energy gap, at least at high fields

So is there a justification for a $\sqrt{B^*}$ dependence? The HLR dimensional
argument can be
rewritten in terms of the effective (CF) parameters. In particular this allows
us to introduce the
cyclotron radius of the composite particles. We write this as
$\l^*_0=\sqrt{\hbar/eB^*}$, which
means we have effectively ignored the presence of all filled CF levels in
determining the inter
(composite) particle spacing. This is a reasonable assumption because the
orthogonality of CF
wavefunctions in different CF LLs means there will only be strong interactions
between CFs of the
same LL. Using this definition we have
\begin{equation}
E^*_c = C^*\frac{e^2}{4\pi\epsilon\epsilon_0 l^*_0} =
C^*\frac{e^2}{4\pi\epsilon\epsilon_0}\sqrt{\frac{eB^*}{\hbar}}
\label{eqn-estar}
\end{equation}
where $C^*$ is now the same constant for all fractions. Using Eq.~(\ref{eqn-e})
to define $M^*$
we can also write
\begin{equation}
M^* = \frac{4\pi\epsilon\epsilon_0}{C^*}
\left(\frac{\hbar}{e}\right)^{3/2}\sqrt{B^*}.
\end{equation}
So in the high field limit this dimensional argument gives a square root
dependence on $B^*$ of both
the gap and the CF mass, with Eq.~(\ref{eqn-estar}) predicting a gapless system
at $\nu=1/2$ as
required. From the fit to Fig.~\ref{fig-mne}b we have an experimental value of
$C^*
=0.063\pm0.005$, valid for all fractions. $C_{\nu}$ has been calculated for
finite
systems\cite{gaptheory} using Laughlin wavefunctions and the hierarchical model
to be
$C_{1/3}$=0.102; $C_{2/5}$=0.063; $C_{3/7}$=0.049 (although there are large
uncertainties
for the higher order fractions). Transforming to CF parameters by $C^* =
\sqrt{2p+1}C_{\nu}$
results in a spread of values of $C^*$=0.177; 0.141; 0.130 respectively for the
1/3, 2/5 and 3/7
fractions. The finite 2DEG thickness will reduce the calculated values by
approximately half and is a
stronger effect at higher field which will reduce this spread.\cite{das}
However, this correction still
leaves a small difference between experiment and theory, which cannot be
attributed to the effects of
level broadening since our measurements of $E^*_c$ are from level center to
center. Future
calculations using wavefunctions appropriate to Composite Fermions may clarify
the situation.

Below about 2T, in the region where the mass has become almost constant, the
energy gap is
smaller than predicted by the strict square root behaviour and by 0.5T there is
almost a factor of 2
difference. This is more obvious in the plot of effective mass
Fig.~\ref{fig-mne}a, where the square
root behaviour is shown as a dotted line. Below 2T the experimental mass values
are considerably
larger than the $\sqrt{B^*}$ behaviour. This could be interpreted as $C^*$
tending towards zero
as $\nu=1/2$ is approached where the dimensional estimation is expected to
break down, as it must
at very low effective fields or $M^*$ would become less than the single
particle mass. Such a
decrease in $C^*$ would have the effect of allowing the gapless system to exist
over a finite range
of field either side of $\nu=1/2$, but would not lead to a divergent mass as
$B^*\rightarrow0$\cite{du,mano} unless $C^*\rightarrow0$ faster than
$\sqrt{B^*}$. Thus the
upshot of this discussion is that while the high effective field data can be
described quite well with a
$\sqrt{B^*}$ dependence, this breaks down as $\nu=1/2$ is approached where the
dimensional
argument fails and corrections to the mean field theory of CFs must be applied.
In the low field
region our original description of the data by Eq.~(\ref{eqn-mass}) remains far
more accurate, while
still adequately describing the high field behaviour, so that from a purely
empirical point of view the
linearly increasing mass remains the most general description.

In conclusion we have measured values for the Composite Fermion effective mass
in pulsed
magnetic fields of up to 50T. $M^*$ shows the same approximately linear
increase with effective
magnetic field that we found earlier at low fields and reaches a value of
$1.6m_e$ for an effective
field of 14T, achieved by placing $\nu=1/3$ at 42T. We correspondingly find
that the CF cyclotron
energy increases sub-linearly with $B^*$ and that a single $\sqrt{B^*}$
behaviour agrees very well
with data from all the fractions with $B^*>2$T. We have discussed some possible
origins of this
behaviour but in the absence of detailed CF energy level calculations can draw
no definitive
conclusions. The results presented here strongly support the Composite Fermion
interpretation of the
FQHE and show that the most significant parameter in determining the properties
of the CF particles
is the effective field $B^*$. This behaviour has only become clear by extending
the study of CF
properties to high densities using pulsed magnetic fields.

MvdB acknowledges support from the EC Human Capital and Mobility programme.

\newpage

\begin{table}
\caption{Sample parameters}
\label{tab:param}
\begin{tabular}{lddd}
Sample  & Spacer Layer & Electron Density &   Mobility \\
   &  \AA  & $10^{15}{\rm m}^{-2}$   &  m$^2$/Vs \\
\hline
G902a &  200 &     4.50    &  160 \\
G902b &  200 &     4.84    &  180 \\
G148   &  400 &     3.26    &  220 \\
G627   &  400 &     3.05    &  370 \\
G650   &  800 &     2.26    &  630 \\
G640   & 1200 &    1.17    &  680 \\
G641   & 1600 &    0.90    &  400 \\
G646   & 2400 &    0.63    &  200
\end{tabular}
\end{table}


\begin{references}
\bibitem{hlrnote} As HLR evaluate $l_0$ at $\nu=1/2$ but use $C_{1/3}$ their
value for $M^*$ is
ambiguous.
\end{references}
\end{document}